\documentclass[11pt]{article}
\usepackage{graphicx}
\usepackage{amsmath}
\usepackage{amssymb}
\usepackage{amsxtra}
\def\s{{\,\rm s}}
\def\g{{\,\rm g}}

\def\MeV{\,{\rm MeV}}
\def\GeV{\,{\rm GeV}}
\def\TeV{\,{\rm TeV}}

\def\({\left(}
\def\){\right)}

\def\beq{\begin{equation}}
\def\eeq{\end{equation}}
\title{Conspiracy of BSM physics and cosmology}
\author{M.Yu. Khlopov\\Institute of Physics, Southern Federal University\\ Stachki 194 Rostov on Don 344090, Russia\\
\and
          National Research Nuclear University MEPhI \\(Moscow Engineering Physics Institute),\\ 115409
Moscow, Russia
\and
          APC laboratory 10, rue Alice Domon et Leonie Duquet\\ 75205 Paris Cedex 13, France, e-mail khlopov@apc.univ-paris.fr}

\begin{document}
\maketitle

\begin{abstract}
The lack of experimental evidence at the LHC for physics beyond the Standard model (BSM) of elementary particles together with necessity of its existence to provide solutions of internal problems of the Standard model (SM) as well as of physical nature of the basic elements of the modern cosmology demonstrates the \textit{conspiracy} of BSM physics. Simultaneously the data of precision cosmology only tighten the constraints on the deviations from the now standard $\Lambda$CDM model and thus exhibit \textit{conspiracy} of the nonstandard cosmological scenarios. We show that studying new physics in combination of its physical, astrophysical and cosmological probes, can not only unveil the conspiracy of BSM physics but will also inevitably reveal nonstandard features in the cosmological scenario.
\end{abstract}



\section{Introduction}\label{intro}
The now standard description of the structure and evolution of the Universe is based on inflationary models with baryosynthesis and dark matter/ energy. The interpretation of the data of precision cosmology ascribes about $95\%$ of the modern cosmological energy density to the impact of physics beyond the Standard model (BSM) of elementary particles. BSM physics is involved in virtually all the mechanisms of inflation and baryosynthesis, explaining the initial conditions of the cosmological evolution. It makes the observed homogeneous and isotropic expanding Universe, origin and structure of its inhomogeneities with their observed baryon asymmetry an evident reflection of the BSM physics.

The problem of experimental studies of BSM physics is generally related with necessity to address effects of a high energy scale $F$ \footnote{(Henceforth,if it is not otherwise specified, we use the units $\hbar=c=k=1$)}. At the energy release $E \ge F$ it leads to appearance of new heavy particles with the mass $M \sim F$ or new interactions that manifest their full strength at these energies. If the energy is much less, than $F$, only virtual effects of new physical scale are possible, which are suppressed by some power of $E/F$. Therefore we can either turn to rare low energy processes, in which new high energy physics phenomena can appear, like proton decay, or probe at the currently available energies $E$ the extensions of the Standard model (SM), which involve new physics at scales $F \le E$. Probes for supersymmetric (SUSY) models at the LHC corresponded to the latter case, but the lack of positive evidence for existence of SUSY particles at the energy of hundreds GeV probably moves the SUSY scale to higher energies, at which direct search of SUSY particle production at the LHC is not possible.

The only experimentally proven evidence for new physics is the effect of neutrino oscillations, but the physical nature of neutrino mass is still unknown.  

Following \cite{me} we characterize here the current situation as the \textit{conspiracy} of the BSM physics: there is no doubt in its existence, but all its features are hidden, since the experimental data put only more and more stringent constraints on the new physics effects. We discuss the physical motivation for extension of SM model and their possible physical, astrophysical and cosmological signatures in Section \ref{bsmp}. We draw attention in the Section \ref{bsmc} that BSM physics involved in the description of the now standard cosmological model (which we consider in Section \ref{bsmp} as the motivation for the SM extension) should inevitably add nonstandard model dependent features like a plethora of non-WIMP forms of dark matter, primordial black holes or antimatter domains in the baryon asymmetrical Universe. We express the hope in the Conclusion (Section \ref{conclusions}) that revealing of specific model dependent signatures of BSM physics can not only unveil its conspiracy, but can also enrich the theory of structure and evolution of the Universe by nonstandard cosmological scenarios.
\section{Motivations for the SM extension}\label{bsmp}
\subsection{Physics of neutrino mass}
The discovery of neutrino oscillations proves the existence of the nonzero mass of neutrino. It may be considered as a manifestation of BSM physics, since neutrinos are strictly massless in the Standard model. However, the very existence of neutrino mass doesn't shed light on its physical nature and the corresponding new physics.

Neutrino mass term relates ordinary left-handed neutrino state to some right handed state. The latter can be ordinary right-handed antineutrino. It corresponds to Majorana mass term, in which lepton number $L$ conservation is violated and $L$ changes as $\Delta L= 2$. In the SM lepton number is conserved at the tree level and Majorana mass term is the example of BSM physics.

Smallness of ordinary neutrino Majorana mass $m_{\nu}$ relative to the Dirac mass $m_D$ of the corresponding charged lepton is explained by "see-saw" mechanism, involving right handed neutrino with large Majorana mass $M$, so that ordinary neutrino mass is given by \beq m_{\nu} = \frac{m_D^2}{M} = \frac{m_D}{M}m_D \ll m_D. \label{seesaw} \eeq Majorana mass term of electron neutrino leads to neutrinoless double beta decay. In the nonrelativistic limit interaction of Majorana neutrino with nuclei is proportional to spin operator acting on nuclear wave function. It leads to spin dependent interaction of nonrelativistic Majorana neutrino with nuclei.

Another possibility is a Dirac neutrino mass term. It corresponds to transition to a new state of sterile right handed neutrino. Such neutrino doesn't participate in the ordinary weak interactions, being another possible example of BSM physics, related to the mechanism of neutrino mass generation. 

In the nonrelativistic limit Dirac neutrino interaction with nuclei is spin independent and leads to coherent scattering of low energy neutrinos in the matter. V.Shwatsman has noted in his diploma work in late 1960s that neutrino with mass $m$ and velocity $v$ can scatter coherently on the piece of matter with size $l \sim \hbar/(mv)$ and cause its acceleration. This idea, published in \cite{ZKrev,shwarts} was probably the first step towards direct detection of cosmological dark matter.

It is the stable prediction of the Big Bang theory that primordial thermal neutrino background should exist with number density \beq n_{\nu \bar \nu} = \frac{3}{11} n_{\gamma}, \label{nun} \eeq where $n_{\gamma} \approx 400 cm^{-3}$ is the the number density of CMB photons. Multiplied by neutrino mass it gives the predicted contribution of relic massive neutrinos to the cosmological density. Experimental constraints on the mass of electron neutrinos (see \cite{katrin} for the latest results) together with the data on the neutrino oscillations exclude explanation of the measured dark matter density by this contribution. However, while ordinary massive neutrinos cannot play dominant dynamical role in the Universe, BSM physics of neutrino mass can lead to important cosmological effects, like sterile neutrino dark matter \cite{sterile}.

\subsection{Supersymmetry and the SM problems}
SUSY models provide natural solution for the internal SM problems, if the SUSY scale is in the range of several hundred GeV. 

Then contribution of SUSY partners in loop diagrams of radiative effects in the Higgs boson mass cancel the quadratic divergent contribution of the corresponding SM particles. Renormalization group analysis of evolution of scalar field potential from superhigh energy scale leads to the Higgs form of this potential at lower energy, explaining the nature of the electroweak symmetry breaking.

R-parity or some continuous symmetry provides stability of the lightest SUSY particle. Such particle with mass of several hundred GeV has interaction cross section at the level of weak interaction and can play the role of Weakly Interacting Massive Particle (WIMP) candidate for dark matter. 

The lack of experimental signatures for SUSY particles at the LHC as well as of positive result of underground WIMP searches \footnote{Though interpretation of positive result of DAMA/NaI and DAMA/LIBRA experiments in the terms of WIMPs is not excluded \cite{DAMAtalk,fabio}, theoretical analysis \cite{kang}, proving such a possibility indicates its contradiction with the results of XENON1T \cite{xenon} and PICO \cite{pico} experiments.} implies nontrivial ways of search for SUSY (see \cite{ellis} for the latest review).

In the extreme case SUSY scale may be close to the scale of Grand Unfication (GUT). This case implies non-SUSY solution for the problem of divergence of the Higgs mass and origin of the electroweak symmetry (see the next subsection), but has the advantage to unify all the four fundamental natural forces, including gravity, in the framework of Supergravity. Starobinsky supergravity can provide simultaneous BSM solution for dark matter in the form of superheavy gravitino \cite{ketovsym,ketov,ketov2} and Starobinsky inflation \cite{star}. This solution can be hardly probed by any direct experimental mean and makes cosmological consequences the unique way for its indirect test.
\subsection{Nonsupersymmetric solutions. Composite Higgs. Multiple charged particles}
Nonsupersymmetric solution for the problem of Higgs mass divergence may be related to the composite nature of Higgs boson \cite{kaplan1,kaplan2,arkani1,arkani2,pomarol1,CTTP}. Then this divergence is cut at the scale, at which Higgs constituents are bound. In parallel such constituents can form bound states with exotic charges. Such situation can take place in the model of composite Higgs based on Walking Technicolor (WTC) \cite{Sannino:2004qp,Hong:2004td,Dietrich:2005jn,Dietrich:2005wk,Gudnason:2006ug,Gudnason:2006yj}.

The minimal walking technicolor model (WTC) involves two techniquarks,$U$ and $D$. They transform
under the adjoint representation of a $SU(2)$ technicolor gauge
group. Neutral techniquark-antitechniquark state is associated with the Higgs boson. Six bosons $UU$, $UD$, $DD$, and their corresponding
antiparticles carry a technibaryon number. If the technibaryon number is conserved, the lightest technibaryon should be stable.

Electric charges of $UU$, $UD$ and $DD$ are given in general by $q+1$, $q$, and $q-1$, respectively, where $q$ is an arbitrary real number \cite{4,KK,KK1}. To compensate the anomalies the model includes in addition
technileptons $\nu'$ and $\zeta$ that are technicolor singlets. Their electric charges are in
terms of $q$, respectively, $(1-3q)/2$ and $(-1-3q)/2$.

Fractional value of $q$ would correspond to stable fractionally charged techniparticles. Their creation in the early Unvierse would lead to their presence in the terrestrial matter that is severely constrained by the experimental data. On the same reason, stable techniparticles should not have odd charge $2n+1$. Positively charged $+(2n+1)$ stable particles are bound with electrons in anomalous isotopes of elements with $Z=2n+1$. Negatively charged particles with charge $-(2n+1)$, created in the early Unvierse, bind with $n+1$ nuclei of primordial helium, produced in the Big Bang Nucleosynthesis, and form a +1 charged ion that binds with electrons in atoms of anomalous hydrogen. The experimental data put severe constraints on such anomalous isotopes.

The case of stable multiple charged particles with even value of negative charge $-2n$ avoids these troubles, since it forms with $n$ nuclei of primordial helium neutral dark atom. Their bound states with primordial helium can play the role of dark matter and can even solve the puzzles of dark matter searches (see \cite{me,iop,front,hadronic} for the latest review).
\subsection{Axion and axion-like models}\label{png}
The popular solution for the problem of strong CP violation in QCD involves the additional $U(1)_{PQ}$ symmetry which provides automatic suppression of the CP-violating $\theta$-term \cite{PQ}. Breaking of this Peccei-Quinn symmetry spontaneously at the scale $f$, followed by its manifest breaking at the scale $\Lambda \ll f$ results in appearance of a pseudo-Nambu-Goldstone (PNG) particle, axion, $a$. In the axion models the second step of breaking is generated by instanton transitions. 

The mass of  axion is given by \cite{sochi} \beq
m_a=C m_{\pi}f_{\pi}/f,\label{axion}\eeq where $m_{\pi}$ and $f_{\pi}\approx m_{\pi}$ are the pion mass and constant, respectively. The constant $C\sim 1$ depends on the choice of the axion model. The relationship (\ref{axion}) of axion to neutral pion makes possible to estimate the cross section of axion interactions from the corresponding cross section of pion processes multiplied by the factor $(f_{\pi}/f)^2$.

The existence of $a \gamma \gamma$ vertex leads to a two-photon decay of axion, as well as to effects of $a \gamma$ conversion \cite{raffelt} like axion-photon conversion in magnetic field (see e.g. \cite{soda} for review and references). The principles of experimental search for axion by "light shining through walls" effects are based on such a conversion \cite{sikivie}.

Axion couplings to nondiagonal quark and lepton transition can lead to rare processes like $K \rightarrow \pi a$ or $\mu \rightarrow e a$. In the gauge model of family symmetry breaking \cite{berezhiani5} the PNG particle called \textit{archion}  shares properties of axion with the ones of singlet Majoron and familon, being related to the mechanism of neutrino mass generation. 

In the axion-like models the condition of Eq. (\ref{axion}) is absent and the mass of the PNG particle may be very small. 

In cosmology, in spite of a very small mass (\ref{axion}) primordial axions appear in the ground state of Bose-Einstein condensate and, being created initially nonrelativistic, represents a specific form of Cold Dark Matter.
\subsection{BSM physics of the standard cosmology}
The now Standard cosmological model involves inflation to explain the homogeneity and isotropy of the Universe as well as initial impulse for Big Bang expansion. Observed absence of antimatter objects is explained by baryosynthesis, in which baryon asymmetry was generated in the intially baryon symmetric Universe. Formation and evolution of Large Scale Structure is described in the framework of the standard $\Lambda$CDM model, assuming dominance in the modern total cosmological density of dark energy with vacuum-like equation of state (cosmological constant $\Lambda$ in the simplest case) and dark matter dominating in the matter content of the Universe. All these elements of the Standard Cosmological model imply BSM physics, making the observational confirmation of this model an evidence for existence of BSM physics.

On the other hand, the data of precision cosmology (planck15,planck18) analysed in the terms of parameters of this standard cosmological model continuously tighten the constraints on deviations of the measured parameters from the model predictions. These measured parameters involve dark matter density $\Omega_{DM} h^2 = 0.120\pm 0.001$, baryon density $\Omega_b h^2 = 0.0224\pm 0.0001$ (where the dimensionless constant $h$ is the modern Hubble constant $H_0$ in the units of 100 km/s/Mpc), scalar spectral index $n_s = 0.965\pm 0.004$, and optical depth $\tau = 0.054\pm 0.007$ \cite{planck18}. These results are only weakly dependent on the cosmological model and remain stable, with somewhat increased errors, in many commonly considered extensions. Assuming the $\Lambda$CDM cosmology, the inferred late-Universe parameters were determined: the Hubble constant $H_0 = (67.4\pm 0.5)$ km/s/Mpc; matter density parameter $\Omega_m = 0.315\pm 0.007$; and matter fluctuation amplitude $\sigma_8 = 0.811\pm 0.006$. Combining with the results of studies of baryon acoustic oscillations (BAO) by measurement of large scale distribution of galaxies \cite{boss} \footnote{Such oscillations were first discussed by A.D. Sakharov \cite{SakhOsc} and are also called {\it Sakharov oscillations}} Planck collaboration has constrained the effective extra relativistic degrees of freedom to be $N_{\rm eff} = 2.99\pm 0.17$, and the sum of neutrino mass was tightly constrained to $\sum m_\nu< 0.12$. These results prove the basic ideas of inflationary model with baryosynthesis and dark matter/energy, but cannot provide definite choice for the corresponding BSM physics. 

 PLANCK collaboration has found no compelling evidence for extensions to the $\Lambda$CDM model, but has mentioned the $3\sigma$ difference with the results of local determination of $H_0$ \cite{riess}. Such a discrepancy may be a hint to a necessity to extend the standard cosmological model.
 
Indeed, the \textit{conspiracy} of Beyond the Standard model (BSM) Cosmology \cite{me} is puzzling taking into account the plethora of nontrivial cosmological consequences of BSM particle models. Some of these nonstandard features which have probably found their experimental evidence are discussed in the next Section \ref{bsmc}. 
\section{Features of BSM cosmology}\label{bsmc}
\subsection{Plethora of dark matter candidates}
Well motivated BSM models offer a plethora of dark matter candidates. In the essence such candidates follow from the extension of the SM symmetry. If the additional symmetry acting on new sets of particles is strict or nearly strict, the lightest particles that possess this symmetry are stable or sufficiently long living to play the role of dark matter. In addition to massive sterile neutrinos, superheavy gravitino or invisible axion that follow respectively from solutions of the origin of neutrino mass, Starobinsky supergravity or solution of the problem of strong CP violation in QCD there are mirror or shadow particles, whose existence is related to restoration of equivalence of left- and right-handed coordinate systems. Grand Unification, string phenomenology or phenomenology of extra dimensions extend this list by many other nontrivial candidates accompanied by a very extensive hidden sector of new particles and fields. Such extensions naturally lead to multicomponent dark matter that can include unstable or decaying components, like it takes place in the model of broken family gauge symmetry \cite{berezhiani5}  (see e.g. \cite{4,dmrps,sochi} for review and references).

In this large list of possibilities the model of dark atoms, in which stable $-2n$ charged particles are bound with $n$ nuclei of primordial helium, is of special interest not only owing to the minimal set of the involved new physics parameters (their number is reduced to the mass of a hypothetical negatively charged stable particles only), but also since it may provide a solution for controversial results of direct dark matter searches.

The idea of this solution is that nuclear interacting dark atoms are slowed down in the terrestrial matter and thus cannot cause significant nuclear recoil in the underground detector. However, in the matter of these detectors dark atoms can bind with intermediate mass nuclei with the binding energy of few keV (see \cite{me,iop,hadronic} for recent review and references). Since the concentration of dark atoms in the matter of underground detectors is adjusted to their incoming cosmic flux, energy release in such binding should experience annual modulations. It explains positive results of DAMA/NaI and DAMA/LIBRA experiments. In a simple rectangular wall and well approximation it was shown in \cite{Levels1} that a level of about 3 keV can exist in binding of dark atoms with intermediate mass nuclei and doesn't exist for heavy nuclei, like xenon, explaining absence of positive results in the corresponding detectors. If such level exists, transition to it is determined by isospin violating electric dipole operator and its rate is proportional to the temperature, being suppressed in cryogenic detectors \cite{me,iop,hadronic}.

The open problem of this explanation is a selfconsistent treatment of Coulomb and nuclear interactions of dark atoms. Such treatment needs special study in the lack of all the usual approximations of atomic physics: there are no small parameters like small ratio of sizes of nucleus and atom and the electroweak interaction of electronic shell. Dark atoms has strongly interacting nuclear shell with the radius of the order or equal to the nuclear radius.

Dark atom cosmology contains such notrivial features as Warmer than Cold Dark Matter scenario and can explain the observed excess of radiation in positron annihilation line from the center of Galaxy as indirect effect of dark atoms (see \cite{me,iop,front,hadronic} for recent review and references). This explanation assuming electron-positron pair production in de-excitation of dark atoms excited in collisions in the center of Galaxy is possible only for a narrow range around 1.25 TeV of the mass of dark atom, which is determined by its  constituent with multiple negative charge \cite{me,sochi}. It challenges search for multiple charged stable particles at the LHC that provides complete test of such an explanation \cite{2cBled}. 

In a two-component dark atom model, a possibility to explain the observed excess of high energy positrons by decays of $+2$ charged dark atom constituents was proposed in \cite{AHEP}. However, any source of positrons is simultaneously the source of gamma radiation and to avoid contradiction with the observed gamma background the mass of the decaying $+2$ consituent of dark atom should be less, than 1 TeV. Moreover, in view of the difference of propagation in the Galaxy by gamma radiation and positrons the condition not to exceed the observed gamma background may cause troubles for any explanation for the high energy positron excess, involving indirect effects of dark matter \cite{kostya}. In any case, the results of searches for stable double charged particles in the ATLAS experiment at the LHC put lower limit on the mass of such particles \cite{Aad2019}, excluding explanation of high energy positron anomaly by decaying $+2$ charged constituents of dark atoms \cite{me,sochi}.
\subsection{Primordial Black holes}
Strong primordial inhomogeneities are a prominent tracer of BSM physics of very early Universe and Primordial Black Holes (PBH) are the most popular example of this kind (see e.g. \cite{ketovsym,PBHrev} for review and references). To form a black hole in the homogeneously expanding Universe the expansion should stop in some region and it corresponds to a very strong inhomogeneity \cite{ZN,hawking1,hawkingCarr}. In the universe with equation of state \begin{equation}
p=\gamma \epsilon \label{EqState} \end{equation} with numerical
factor $\gamma$ being in the range \begin{equation}
\label{FacState}0 \le \gamma \le 1\end{equation} the probability of
forming a black hole from fluctuations within the cosmological horizon is
given by \cite{carr75}
\begin{equation}
W_{PBH} \propto \exp \left(-\frac{\gamma^2}{2
\left\langle \delta^2 \right\rangle}\right),
\label{ProbBH}
\end{equation}
where $\left\langle \delta^2 \right\rangle \ll 1$ is the amplitude of density fluctuations. For relativistic equation of state ($\gamma = 1/3$) the probability (\ref{ProbBH}) is exponentially small. It can increase, if the amplitude of density fluctuations in the early Universe was much larger, than in the period of galaxy formation, or the equation of state was much softer, corresponding to matter dominated stage with $\gamma = 0$.  

Therefore PBH origin may be related with early matter dominated stages, phase transitions in the early Universe or nonflat features in the spectrum of primordial density fluctuations. All these phenomena are not only originated from BSM physics, but also represent strong deviation from the Standard cosmological scenario.

PBHs with mass $M \le 10^{15} \g$ evaporate by the mechanism of Hawking \cite{hawking4,21}. This process is the universal process of production of any type of particles with mass 
$$m \le T_{evap} \approx 10^{13} \GeV\frac{1\g}{M}.$$
It can be the source of superweakly interacting particles, like gravitino \cite{KBgrain} as well as of fluxes of particles with energy much larger, than the thermal energy of particles in the surrounding medium. It causes non equilibrium processes in the hot Big Bang Universe, nonequilibrium cosmological nucleosynthesis \cite{khlopov31}, in particular.  

PBHs with mass $M \ge 10^{15} \g$ should survive to the present time and represent a specific form of dark matter. It was noticed in \cite{pbhClusters} that taking into account PBH formation in clusters the constraints on PBH contribution into the total density \cite{CarrKeun} can be relaxed and even the possibility of PBH dominant dark matter is not excluded. It would make primordial nonhomogeneities in the form of PBHs the dominant matter content of the modern nonhomogeneities.

Mechanism of PBH cluster formation can be illustrated with the use of the axion-like model, discussed in subsection \ref{png}, in which the first step of symmetry breaking at scale $f$ takes place on the inflationary stage \cite{sochi,PBHrev}. Then at the second stage of the symmetry breaking at $T \sim \Lambda$ closed massive walls are formed so that the larger wall is accompanied by a set of smaller walls. Their collapse form a PBH cluster, in which the range of PBH masses $M$ is determined by the model parameters $f$ and $\Lambda$ \cite{sochi,book2}
\beq
\label{Mmaxmin} f(\frac{m_{pl}}{\Lambda})^2 \le M \le
\frac{m_{pl}}{f}m_{pl}(\frac{m_{pl}}{\Lambda})^2\eeq 
Here the minimal mass is determined by the condition that the width of wall doesn't exceed its gravitational radius, while the upper limit comes from the condition that the wall enters horizon, before it starts to dominate within it \cite{book2}.
At $\Lambda < 100 \MeV (m_{pl}/f)^{1/2}$ the maximal mass exceeds 100$M_{odot}$. Collapse of massive walls to such black holes takes place at
\beq t>\frac{m_{pl}}{f}\frac{m_{pl}}{\Lambda^2}. \eeq At $\Lambda < 1 \GeV$ and $f=10^{14} \GeV$ it happens at $t>0.1 \s$, what can lead to interesting observable consequences.

Closed wall collapse leads to primordial gravitational wave (GW) 
 spectrum, estimated as peaked at \cite{sochi}\beq
\label{nupeak}\nu_0=3 \times 10^{11}(\Lambda/f){\rm Hz}. \eeq Their estimated contribution to the total density can reach \beq \label{OmGW}\Omega_{GW} \approx 10^{-4}(f/m_{pl}),\eeq 
being at $f \sim 10^{14}$ GeV $\Omega_{GW}\approx
10^{-9}.$ For $
1<\Lambda<10^8\,{\rm GeV}$ the maximum of the spectrum corresponds to
\beq 3 \times 10^{-3}<\nu_0<3 \times  10^{5}\,{\rm Hz},\eeq 
being in the range from tens to 
thousands of Hz a challenge for  LIGO/VIRGO grvitational wave searches.

Predictions for Gravitational wave signals from PBH coalescence in cluster involve study of cluster evolution, which is now under way \cite{pbhClusters}.

Being in cluster, PBHs with the masses of tens $M_{\odot}$ form binaries much easier, than in the case of their random distribution, as well as formation of such PBHs in collapse of first stars is rather problematic. In this aspect detection of signals from binary BH coalescence in the gravitational wave experiments \cite{gw1,gw2,gw3,gw4,gw5} may be considered as a positive evidence for this scenario \cite{sochi}. Repeatedly detected signals localized in the same place would provide successive support in its favor or exclusion \cite{sochi,pbhClusters,Bringmann}. The~existing statistics is evidently not sufficient to make any definite conclusion on this possibility. However, repeating detection of four GW signals in the August of 2017 noted in GWTC catalog \cite{GWTC} may be an interesting hint to such a possibility \cite{me,sochi}. 

Primordial black holes reflect strong inhomogeneity of the very early Universe. Their production in a significant amount is not a necessary consequence of all the models of very early Universe. However, it is just this model dependent character provides a very sensitive probe of BSM physics and the confirmation of PBH existence can severely tighten the class of possible realistic BSM models. 

The same is true for the existence of antimatter objects in baryon asymmetric Universe, which can reflect strong nonhomogeneity of the baryosynthesis.  
\subsection{Antimatter and Baryon Asymmetry}
The baryon asymmetry of the Universe is related with the evident dominance of matter over antimatter in the visible part of the Universe. The set of astrophysical data puts only constraints on the possible amount of macroscopic antimatter.
However severe, these constraints still don't exclude completely the existence of antimatter objects, which can be formed in antimatter domains in baryon asymmetric Universe originated from the strongly nonhomogeneous baryosynthesis \cite{CKSZ,Dolgov2,GC,ANTIHE,KRS2,Dolgov,Dolgov3} (see \cite{book2,ketovsym,Dolgov} for review and references). 

If created, antimatter domains should survive in the surrounding matter to the present time. It puts a lower limit on its size being in terms of its mass about $10^3 M_{odot}$ \cite{GC,ANTIHE,KRS2} that corresponds to a minimal mass of globular clusters. If antimatter globular cluster is formed in our Galaxy, it may be the source of cosmic ray antinuclei.

However exotic, the hypothesis on antimatter globular cluster in our Galaxy \cite{GC} doesn't contradict observations, if the mass of the cluster doesn't exceed the limit \beq M \le 10^5 M_{odot}. \label{agc} \eeq Indeed, globular clusters are an old population of the Galaxy being dominantly in halo, where matter gas density is low. Their gravitational potentials are not sufficient to hold matter, lost by stars by stellar winds or supernova explasions. In the case of antimatter cluster, it means that there is no antimatter gas within it and matter gas that enters the cluster annihilate only on antimatter stellar surfaces. Taking into account low density of matter gas in halo and relatively small surface on which it can annihilate, one can conclude that antimatter globular cluster is expected to be a rather faint gamma ray source. The upper limit (\ref{agc}) follows from the condition that the antimatter lost by antimatter stars and polluting the Galaxy doesn't cause overproduction of gamma ray background from annihilation with the matter gas \cite{GC,ANTIHE,KRS2}.

It was noted in \cite{GC,ANTIHE,KRS2} that cosmic antihelium flux may be a profound signature for an antimatter globular cluster in our Galaxy. Symmetry in physics of matter and antimatter would make antihelium-4 the second by abundance element of antimatter. In addition to antihelium lost by antimatter stars its cosmic fluxes can increase due to destruction of heavier antinuclei in their annihilation with matter. Rough estimation of the expected antihelium flux as simply proportional to the ratio of the mass of antimatter cluster to the total mass of the Galaxy predicts that it should be within the reach by AMS02 experiment to 2024.

This prospect makes necessary to specify the predictions for the cosmic antihelium flux in more details and such analysis can be based on our knowledge of properties of globular clusters. However, one should take into account that antimatter stars may have properties much different from ordinary stars with correspondingly different observational signatures \cite{Blinnikov:2014nea}.

Possible detection of cosmic antihelium-3 nuclei by AMS02 experiment together with some detected events that may correspond to antihelium-4 cannot find natural astrophysical explanation \cite{poulin} and may be strong evidence for existence of macroscopic forms of antimatter in our Galaxy.
\section{Conclusions}\label{conclusions}
Plethora of BSM physics involves, pending on the particular model, various combinations of its physical, astophysical and cosmological signatures. Such model dependent predictions lead to nontrivial cosmological features that can provide potentially observable deviations from the predictions of the standard cosmological model.

We have drawn special attention to some, at first glance exotic, predictions, like nuclear interacting dark atoms, massive PBH clusters or antimatter stars in our Galaxy, since they can explain the corresponding experimental anomalies, such as the puzzles of direct dark matter searches, origin of coalescensing massive black holes or experimental evidence for cosmic antihelium. If these explanations are confirmed, they strongly tighten the class of viable BSM models and add the corresponding nonstandard features to the cosmological scenario. Reminding Ya.B.Zeldovich, we can repeat after him that "though the probability for existence of these phenomena seems low, the expectation value of their discovery can be hardly overestimated".
\section*{Acknowledgements}
I am grateful to MDPI journal Symmetry for travel support at the XXII Bled Workshop. The work was supported by grant of Russian Science Foundation (Project No-18-12-00213).


\end{document}